\documentclass[seceq]{ptptex}
\usepackage{wrapft}
\usepackage{graphicx}

\def\beq{\begin{eqnarray}}
\def\eeq{\end{eqnarray}}
\def\bsub{\begin{subequations}}
\def\esub{\end{subequations}}
\def\b{\begin{equation}}
\def\bs{\begin{split}}
\def\es{\end{split}}
\def\e{\end{equation}}

\begin{document}

\title{A possibility of quark spin polarized phase\\
in high density quark matter}

\author{Yasuhiko {\sc Tsue}$^{1,2}$, {Jo\~ao da {\sc Provid\^encia}}$^{2}$, {Constan\c{c}a {\sc Provid\^encia}}$^{2}$, \\
{Masatoshi {\sc Yamamura}}$^{3}$ and {Henrik {\sc Bohr}}$^{4}$
}

\inst{$^{1}${Physics Division, Faculty of Science, Kochi University, Kochi 780-8520, Japan}\\
$^{2}${Center for Computational Physics, Departamento de F\'{i}sica, Universidade de Coimbra, 3004-516 Coimbra, 
Portugal}\\
$^{3}${Department of Pure and Applied Physics, 
Faculty of Engineering Science, Kansai University, Suita 564-8680, Japan}\\
$^{4}${Department of Physics, B.307, Danish Technical University, DK-2800 Lyngby, Denmark}
}

\abst{
It is shown that the quark spin polarization may occur for each quark flavor by the use of the Nambu-Jona-Lasinio (NJL) model 
with a tensor-type four-point interaction between quarks, while 
the two-flavor color superconducting (2SC) phase in two-flavor case 
may be realized at high density quark matter.
}


\maketitle

\section{Introduction}

One of recent interests in the physics governed by the quantum chromodynamics (QCD) may 
be to clarify the possibility of various phases under extreme conditions such as high temperature, 
high density, strong magnetic field and so on.
Especially, in the region of high baryon density and low temperature, 
which phase exists is one of central problems, which is related to the understanding 
of inside of compact stars. 
For example, it is expected that the color superconducting phase is realized in quark matter. 
Here, another possibility with respect to the realized phase is investigated in the 
region of high baryon density and zero temperature in the context of the physics of compact stars. 
Namely, it is indicated that the quark spin polarization for each quark flavor 
may occur by the use of the NJL model 
with tensor-type four-point interaction between quarks \cite{1,2,3}.

\section{Two-flavor quark spin polarization versus two-flavor color superconductivity}

Let us start with the following Lagrangian density: 
\beq\label{1}
& &{\cal L}={\bar \psi}i\gamma^\mu\partial_\mu\psi+G_S(({\bar \psi}\psi)^2+({\bar \psi}i\gamma_5{\vec \tau}\psi)^2)+{\cal L}_T+{\cal L}_c\ , \nonumber\\
& &\ {\cal L}_T=-\frac{G}{4}\left(({\bar \psi}\gamma^{\mu}\gamma^{\nu}{\vec \tau}\psi)
({\bar \psi}\gamma_{\mu}\gamma_{\nu}{\vec \tau}\psi)
+({\bar \psi}i\gamma_5\gamma^{\mu}\gamma^{\nu}\psi)
({\bar \psi}i\gamma_5\gamma_{\mu}\gamma_{\nu}\psi)\right)\ , \nonumber\\
& &\ {\cal L}_c=\frac{G_c}{2}\sum_{A=2,5,7}\left(
({\bar \psi}i\gamma_5\tau_2\lambda_A\psi^C)(
{\bar \psi}^Ci\gamma_5\tau_2\lambda_A\psi)+
({\bar \psi}\tau_2\lambda_A\psi^C)({\bar \psi}^C\tau_2\lambda_A\psi)\right)\ . 
\eeq
Here, $\psi^C=C{\bar \psi}^T$ with $C=i\gamma^2\gamma^0$ being the charge conjugation operator. 
Also, $\tau_2$ is the second component of the Pauli matrices representing 
the isospin $su(2)$-generator and 
$\lambda_A$ are the antisymmetric Gell-Mann matrices representing the color $su(3)_c$-generator.  
Here, we adopt the mean field approximation in which operators $({\bar \psi}\Gamma^A \psi)({\bar \psi}\Gamma_A \psi)$ are replaced into 
$2\langle{\bar \psi}\Gamma^A \psi\rangle({\bar \psi}\Gamma_A \psi)-\langle{\bar \psi}\Gamma^A \psi\rangle^2$ where 
$\langle \cdots \rangle$ represent the mean-field values. 
We concentrate on quark matter at high baryon density where the chiral symmetry 
is restored in the density region considered here, which leads to $\langle {\bar \psi}\psi \rangle =0$. 
Further, we expand the quark field by using good helicity states. 
After that, we introduce the BCS state. 
As a result, we can derive the thermodynamic potential $\Phi$ as follows: 
\beq\label{2}
& &\Phi(\Delta,F,\mu)
=
2\cdot\frac{1}{V}\sum_{{\mib p}\eta (\varepsilon_{\mib p}^{(\eta)}\leq \mu)}
\left[2(\varepsilon_{\mib p}^{(\eta)}-\mu)-\sqrt{(\varepsilon_{\mib p}^{(\eta)}-\mu)^2+3\Delta^2f_p(\eta)^2}
\right]\nonumber\\
& &\qquad\qquad\quad
+2\cdot\frac{1}{V}\sum_{{\mib p}\eta (\varepsilon_{\mib p}^{(\eta)}> \mu)}^{\Lambda}
\left[(\varepsilon_{\mib p}^{(\eta)}-\mu)-\sqrt{(\varepsilon_{\mib p}^{(\eta)}-\mu)^2+3\Delta^2f_p(\eta)^2}
\right]\nonumber\\
& &\qquad\qquad\quad
+\frac{F^2}{2G}+\frac{3\Delta^2}{2G_c} \ , \quad\\
& &\qquad
F=-G\langle {\bar \psi}\Sigma_3 \tau_3 \psi\rangle, \quad
\Sigma_3=-i\gamma^1\gamma^2=
\left(
\begin{array}{cc}
\sigma_3 & 0 \\
0 & \sigma_3 
\end{array}\right), 
\nonumber\\
& &\qquad
\Delta_A=\Delta_A^*=-G_c\langle{\bar \psi}^Ci\gamma_5\tau_2\lambda_A\psi\rangle ,
\nonumber\\
& &\qquad
f_p(\eta)=\frac{p+\eta e}{\varepsilon_{\mib p}^{(\eta)}}\ , \qquad p=\sqrt{p_1^2+p_2^2+p_3^2}\ , \qquad e=F\frac{\sqrt{p_1^2+p_2^2}}{p}\ , \nonumber 
\eeq
where $\varepsilon_{\mib p}^{(\eta)}=
\sqrt{p_3^2+\left(F+\eta\sqrt{p_1^2+p_2^2}\right)^2}$ represents the single quark energy, and $\mu$, $\eta=\pm$ and $V$ represent the chemical potential, 
the helicity and the volume, respectively. 
Here, $F$ and $\Delta=\Delta_2=\Delta_5=\Delta_7$ represent the quark spin polarized condensate (spin alignment) 
and the color superconducting gap, respectively.~\!\!\footnote{
In \cite{2}, an approximation that $f_p(\eta)=1$ is adopted. Here, we retain the factor $f_p(\eta)$. 
The qualitative behavior of the thermodynamic potential is not changed, but quantitatively, 
the thermodynamic potential is slightly modified and the behavior of the thermodynamic potential becomes smoother.
} 
\begin{figure}[t]
\centering
{\includegraphics[angle=0,width=12.8cm]{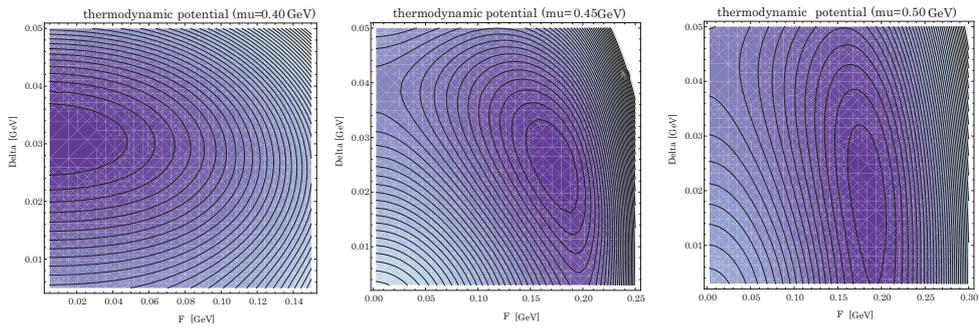}}
\caption{The contour map of thermodynamic potential of each value of chemical potential $\mu$ 
is shown. 
The vertical and horizontal axis represent the color superconducting gap $\Delta$ and the spin polarized condensate $F$, respectively. 
}
    \end{figure}
Figure 1 shows the contour map of the thermodynamic potential.
The parameters used here are $G=20.0$ GeV, $G_c=6.6$ GeV and the three-momentum cutoff $\Lambda=0.631$ GeV, respectively.    
In the region with small chemical potential such as $\mu=0.40$ GeV, the thermodynamic potential has the minimum point 
with $F=0$ and $\Delta \neq 0$, 
which is nothing but the 2SC phase. 
As the chemical potential is increasing, the point with $F\neq 0$ and $\Delta \neq 0$ gives a minimum point. 
Finally, in the region with large chemical potential such as $\mu=0.50$ GeV, the Thermodynamic potential reveals minimum 
with $F\neq 0$ and $\Delta=0$, which is identical with the spin polarized phase. 
Thus, the phase transition from 2SC to SP phases occurs.

\section*{Acknowledgment}
One of the authors (Y.T.) would likes to express his thanks to Mr. H. Matsuoka for his help in the numerical calculation.


\begin{thebibliography}{999}
\bibitem{1}
Y. Tsue, J. da Provid\^encia, C. Provid\^encia and M. Yamamura, Prog. Theor. Phys. 2012, 128, 507.

\bibitem{2}
Y. Tsue, J. da Provid\^encia, C. Provid\^encia, M. Yamamura and H. Bohr, Prog. Theor. Exp. Phys. 2013, 103D01.

\bibitem{3}
Y. Tsue, J. da Provid\^encia, C. Provid\^encia, M. Yamamura and H. Bohr, Prog. Theor. Exp. Phys. 2015, 013D02.

\end{thebibliography}
\end{document}